\documentstyle[11pt,epsfig]{article}

\title{Sensitivity of deexcitation energies of superdeformed secondary minima
to the density dependence of symmetry energy with the relativistic
mean-field theory}
\author{ W. Z. Jiang$^{1,4}$,  Z. Z. Ren$^{2,4}$, Z. Q. Sheng$^2$,
and Z. Y. Zhu$^{3,4}$\\
\small\it $^1$ Department of Physics, Southeast University,\\
\small\it Nanjing 211189, China \\
\small\it $^2$ Department of Physics, Nanjing University, \\
\small\it Nanjing 210093,   China\\
\small\it $^3$ Institute of Applied Physics,\\
\small\it  Chinese Academy of Sciences, Shanghai 201800, China\\
\small\it   $^4$ Center of Theoretical Nuclear Physics,\\
\small\it  National  Laboratory of Heavy Ion Accelerator, Lanzhou
730000,China\\}

\date{}
\bigskip
\begin{document}

\maketitle \baselineskip 21.6pt

\begin{abstract}
\baselineskip 18.0pt The relationship between deexcitation energies
of superdeformed secondary minima relative to ground states and the
density dependence of the symmetry energy is investigated for heavy
nuclei using the relativistic mean field (RMF) model. It is shown
that the deexcitation energies of superdeformed secondary minima are
sensitive to differences in the symmetry energy that are mimicked by
the isoscalar-isovector coupling included in the model. With
deliberate investigations on a few Hg isotopes that have data of
deexcitation energies, we find that the description for the
deexcitation energies can be improved due to the softening of the
symmetry energy. Further, we have investigated deexcitation energies
of odd-odd heavy nuclei  that are nearly independent of pairing
correlations, and have discussed the possible extraction of the
constraint on the density dependence of the symmetry energy with the
measurement of deexcitation energies of these nuclei.

\end{abstract}

\baselineskip 16pt
 \thanks{PACS: 21.65.Ef, 21.10.-k, 21.60.Jz}

\thanks{Keywords: Symmetry energy,  Superdeformation, Relativistic mean-field
models }

\section{Introduction}

The nuclear symmetry energy plays an important role in
astrophysics~\cite{lat01,ho01,ho02,steiner05a}, the structure of
neutron- or proton-rich nuclei, and the reaction dynamics of
heavy-ion collisions, see, e.g., Refs.~\cite{todd,ji05,ba08}.
However, the density dependence of the symmetry energy is still
poorly known, for instance, see Ref.~\cite{ba08}. Recently,
considerable progress has been made in constraining the density
dependence of the symmetry energy using data from heavy-ion
reactions~\cite{ts04,ch05,li05,Fopi07,xiao08}. On the other hand, it
is promising to constrain the density dependence of the symmetry
energy at subsaturation densities by accurately measuring the neutron
skin thickness in heavy nuclei. In the past, Horowitz et. al.
proposed to measure the neutron radius of $^{208}$Pb by virtue of the
parity-violating electron scattering on the neutrons that promises a
1\% accuracy~\cite{ho01,todd,piek06,jeff}. While the precision
measurement is still in progress, it is valuable to explore whether
some other structural properties of finite nuclei are sensitive to
differences in the symmetry energy.

As one knows, the neutron skin thickness in heavy nuclei depends
sensitively on the density dependence of the symmetry energy. The
sensitive probe to differences in the symmetry energy may thus
possibly exist in  systems that undergo a relative variation of the
proton and neutron matter distributions. This relative variation can
follow from the collective excitation or deexcitation between the
ground state and superdeformed secondary minimum (SSM) in heavy
nuclei with appreciable neutron excesses. Especially, the large
relative variation can be expected for nuclei in the  mass region
$A\sim190$ where the prolate superdeformation of the secondary minima
usually occurs with the oblate ground
states~\cite{ma86,dr88,wo92,he98}. In deed, quite different
deexcitation energies of the SSM for nuclei in the region $A\sim190$
were predicted by various models that usually diversify the symmetry
energies, see Ref.~\cite{lau00} and references therein. Since in
obtaining the deexcitation energy the isoscalar ingredients of the
ground state and SSM cancel largely, the variation of the
deexcitation energy can be mainly attributed to the uncertainty of
the symmetry energy that is controlled by the isovector potential.
However, a direct relationship between the deexcitation energy and
the density dependence of the symmetry energy is not available in the
literature. In this work, it is thus meaningful to establish this
relationship. It is the aim of this work to constrain the density
dependence of the symmetry energy in the relativistic mean-field
(RMF) model through deexcitation energies measured and to be measured
for nuclei in the region $A\sim190$. We note that the constraint on
differences in the symmetry energy has been investigated using
properties of the isovector giant  and  pigmy dipole
resonances~\cite{vr03,pa05,pi06,liang07}. Apart from these dynamical
resonances,  the SSM features a static structure.  Indeed, this is an
attempt to constrain the density dependence of the symmetry energy
using the information of atomic masses, since the deexcitation energy
is the difference between masses of the  SSM and ground state.

The paper is organized as follows. In Section \ref{RMF}, we briefly
introduce the formalism of the deformed RMF model for finite nuclei.
Results on the SSM and ground states  of finite nuclei, especially
the deexcitation energies are presented in Section \ref{results}. A
summary is finally given in Section \ref{summary}.

\section{Formalism}
\label{RMF}

 The model lagrangian is written as:
 \begin{eqnarray}
 {\cal L}&=&
{\overline\psi}[i\gamma_{\mu}\partial^{\mu}-M_N+g_{\sigma}\sigma-g_{\omega}
\gamma_{\mu}\omega^{\mu}-g_\rho\gamma_\mu \tau_3 b_0^\mu
  -e\frac{1}{2}(1+\tau_3)\gamma_\mu A^\mu]\psi\nonumber\\
      &  &
    - \frac{1}{4}F_{\mu\nu}F^{\mu\nu}+
      \frac{1}{2}m_{\omega}^{2}\omega_{\mu}\omega^{\mu}
    - \frac{1}{4}B_{\mu\nu} B^{\mu\nu}+
      \frac{1}{2}m_{\rho}^{2} b_{0\mu} b_0^{\mu}-\frac{1}{4}A_{\mu\nu}
      A^{\mu\nu}\nonumber\\
&& +
\frac{1}{2}(\partial_{\mu}\sigma\partial^{\mu}\sigma-m_{\sigma}^{2}\sigma^{2})
+U(\sigma,\omega^\mu, b_0^\mu), \label{eq:lag1}
  \end{eqnarray}
 where $\psi,\sigma,\omega$, and $b_0$ are the fields of
the nucleon,  scalar, vector, and neutral isovector-vector,  with
their masses $M_N, m_\sigma,m_\omega$, and $m_\rho$, respectively.
$A_\mu$ is the photon field. $g_i(i=\sigma,\omega,\rho)$  are the
corresponding meson-nucleon couplings. $F_{\mu\nu}$, $ B_{\mu\nu}$
and $A_{\mu\nu}$ are the strength tensors of $\omega$ and $\rho$
mesons, and photon, respectively
\begin{equation}\label{strength} F_{\mu\nu}=\partial_\mu
\omega_\nu -\partial_\nu \omega_\mu,\hbox{  } B_{\mu\nu}=\partial_\mu
b_{0\nu} -\partial_\nu b_{0\mu},\hbox{ } A_{\mu\nu}=\partial_\mu
A_{\nu} -\partial_\nu A_{\mu}.
\end{equation}
The self-interacting terms of $\sigma$, $\omega$ mesons and  the
isoscalar-isovector coupling  are given generally as
 \begin{eqnarray}
 U(\sigma,\omega^\mu, b_0^\mu)&=&-\frac{1}{3}g_2\sigma^3-\frac{1}{4}g_3\sigma^4
 +\frac{1}{4}c_3(\omega_\mu\omega^\mu)^2,\nonumber\\
 &&+4g^2_\rho  g_\omega^2 \Lambda_{\rm v}
 \omega_\mu\omega^\mu b_{0\mu}b_0^\mu.
 \end{eqnarray}
Here, the isoscalar-isovector coupling term is introduced to modify
the density dependence of the symmetry energy. In fact, the symmetry
energy can be modified by many theoretical factors, for instance, the
isoscalar-isovector coupling terms\cite{ho01}, the isovector-scalar
mesons, the density dependent coupling constants\cite{ma02}, and the
model chirality constraint~\cite{ji07}. In the past, the density
dependence of the symmetry energy had been extensively explored
through the inclusion of the isoscalar-isovector coupling
terms~\cite{ho01,ho02,todd,piek06} in the RMF theory, and this
allowed one modify the neutron skin of heavy nuclei without
compromising the success in reproducing a variety of ground-state
properties~\cite{ho01,todd}. In this work,  the isoscalar-isovector
coupling term is thus  included in the deformed RMF model to modify
the density dependence of the symmetry energy.

Using the Euler-Lagrangian equation, the Dirac equation of motion in
RMF is written as
 \begin{equation} [-i{\bf \alpha}\cdot\nabla +\beta M^*_N+
g_\omega\omega_0({\bf r})+g_\rho\tau_3
 b_0({\bf r})+e\frac{1}{2}(1+\tau_3) A_0({\bf r})]\psi_i({\bf r})=E_\alpha\psi_i({\bf r}),
 \label{eqr} \end{equation} with
$M^*_N=M_N- g_\sigma\sigma({\bf r})$  and $E_\alpha$ being the
nucleon eigen energy. For simplicity, the isospin subscript for the
$\rho$-meson field is omitted hereafter. For the mesons and photon,
the equations of motion are given as
 \begin{equation} \label{eqmeson}
  (\Delta-m_\phi^2)\phi({\bf r}) =
  -s_\phi({\bf r})
  \end{equation}
where for the photon, $m_\phi=0$, and
 \begin{equation} s_\phi({\bf r})=\left\{\begin{array}{cl}
g_\sigma\rho_s({\bf r})-g_2\sigma^2({\bf r})-g_3\sigma^3({\bf r}), &
 \hbox{ $\sigma$ },\\\\
 g_\omega\rho_B({\bf r})-c_3\omega_0^3-8g^2_\omega g^2_\rho\Lambda_{\rm v}\omega_0({\bf r}) b_0^2({\bf r}), &
 \hbox{ $\omega$ },\\\\
g_\rho\rho_3({\bf r})-8g^2_\rho g^2_\omega\Lambda_{\rm v} b_0({\bf
r}) \omega_0^2({\bf r}), &
 \hbox{ rho },\\\\
  e\rho_c({\bf r}),&
\hbox{ photon.}
\\ \end{array}\right.
 \end{equation}
Here $\rho_s$, $\rho_B$, $\rho_3$ and $\rho_c$ are the scalar,
vector, isovector and charge densities, respectively. The total
binding energy is written as
 \begin{eqnarray} \label{eqeng}
E_{total}&=&E_{N}+E_\sigma+E_{\omega_0}+E_{b_0}+E_c
+E_{CM}\nonumber\\
 &=&\sum_\alpha (E_\alpha-M_N) -\frac{1}{2}\int d^3r [g_\sigma\sigma({\bf r})\rho_s({\bf r})+
 \frac{1}{3}g_2\sigma^3({\bf r})+\frac{1}{2}g_3\sigma^4({\bf r}) ]\nonumber\\
 && +
 \frac{1}{2}\int
 d^3r[g_\omega\omega_0({\bf r})\rho_B({\bf r})+\frac{c_3}{2}\omega_0^4({\bf r})]
 \nonumber\\
&&+ \frac{1}{2}g_\rho\int d^3rb_0({\bf r})[\rho_3({\bf r})+8g_\rho
g_\omega^2\Lambda_{\rm v}
 \omega_0^2({\bf r})b_0({\bf r})]  \nonumber\\
  &&+ \frac{1}{2} e\int d^3rA_0({\bf r})\rho_c({\bf r})-\frac{3}{4}41A^{1/3}.
 \end{eqnarray}

In practical calculations, we also include the BCS pairing
interaction using the constant pairing gaps which are obtained from
the prescription of M\"oller and Nix \cite{mn2}:
$\Delta_n=4.8/N^{1/3},\hbox{ } \Delta_p=4.8/Z^{1/3}$ with N and Z the
neutron and proton numbers, respectively.  Here, the cut-off
$82A^{-1/3}$ MeV above the nucleon chemical potentials is used to
normalize the pairing energy. We note that the  BCS description for
nucleon pairing is not as popular as the Bogoliubov approach, e.g.,
see Refs.~\cite{ri96,meng06} and references therein. However, in this
work we do not pursue a complete description of nucleon pairings but
investigate the relative change in  binding energies of the ground
state and SSM with respect to the density dependence of the symmetry
energy. For both ground states and SSM, the numbers of oscillator
shells $N_F=N_B=18$ are used in the basis expansion. The solution of
the coupled Dirac and meson equations can be obtained in an iterative
procedure that can be easily found in the literature~\cite{ga90}, and
it is not reiterated here.

\section{Results and discussions}
\label{results}

In this work, we make analyses based on the calculations with the RMF
parameter set NL3*~\cite{nl3new} that is an improved version of the
parameter set NL3~\cite{nl3}. Also, the isoscalar-isovector coupling
is included to modify the density dependence of the symmetry energy.
This new parameter set has been  successfully tested by the
properties of the ground states and collective excitations of fintie
nuclei. Based on this parameter set, here we investigate the
deexcitation energies relative to the ground states for Hg and Au
isotopes, while attention will be paid to the sensitivity of the
deexcitation energy to differences in the symmetry energy.

Prior to practical calculations, it is useful to write here the
explicit expression of the symmetry energy in the RMF models
\begin{equation}\label{eqsym}
    E_{sym}=\frac{1}{2}\left(\frac{g_\rho}{m_\rho^*}\right)^2 \rho_B
    +\frac{k_F^2}{6E_F^*}=\frac{1}{2\delta}g_\rho b_0
    +\frac{k_F^2}{6E_F^*},
\end{equation}
where $m_\rho^*$ is the $\rho$-meson effective mass with
$m^*_\rho=\sqrt{m_\rho^2+8\Lambda_{\rm v}(g_\omega
g_\rho\omega_0)^2}$, $\delta$ is the isospin asymmetry  with
$\delta=\rho_3/\rho_B$, and $E_F^*$ is the Fermi energy. The first
term is the potential part of the symmetry energy, and the second
term is the kinetic part. The modification to the symmetry energy is
dictated by the potential part through the isoscalar-isovector
coupling.

\begin{figure}[thb]
\begin{center}
\vspace*{-15mm}
\includegraphics[height=10.0cm,width=10.0cm]{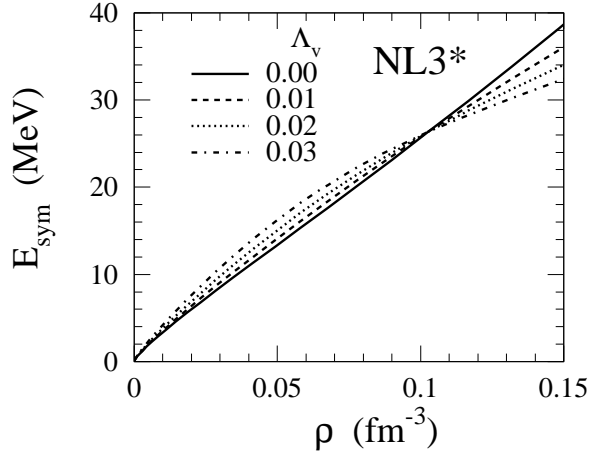}
 \end{center}
\caption{Symmetry energy as a function of the density with the NL3*.
The various $\Lambda_{\rm v}$ are denoted for corresponding
curves.\label{figsym}}
\end{figure}

Since the symmetry energy is not well constrained at saturation
density, some average of the symmetry energy at saturation density
and the surface energy is needed according to the constraint from the
binding energy of nuclei. For a given $\Lambda_{\rm v}$, we follow
Ref.~\cite{ho01} to readjust the $\rho NN$ coupling constant $g_\rho$
so as to keep the symmetry energy unchanged at $k_F=1.15$ fm$^{-1}$.
In doing so, the symmetry energy is softened by the
isoscalar-isovector coupling, as shown in Fig.~\ref{figsym}. As a
result, it was found that the neutron skin thickness relies
sensitively on the $\Lambda_{\rm v}$, while the total binding energy
of $^{208}$Pb just changes by a few MeV with the $\Lambda_{\rm v}$ of
interest~\cite{ho01}. The small change in the total binding energy
results from two cancellation mechanisms. The first mechanism is
inherent in the RMF theory due to the cancellation between the big
scalar attraction provided by the $\sigma$ meson and the big vector
repulsion provided by the $\omega$ meson~\cite{se86}. Though the
isoscalar-isovector coupling modifies predominantly the isovector
potential, its influence can be passed on to the isoscalar potential
through modified nucleon matter distributions. Here, the $\sigma$
field energy changes  with respect to $\Lambda_{\rm v}$ coherently
with the $\omega$ field energy but with an opposite sign.  The second
mechanism can be understood according to the virial theorem, which
can be written in the non-relativistic form as:
\begin{equation}\label{eqviri}
   <\psi_i|\frac{{\bf p}^2}{2M_N}|\psi_i>=\frac{1}{2}
   <\psi_i|{\bf r}\cdot\nabla V({\bf r})|\psi_i>,
\end{equation}
where $V({\bf r})$ is the nuclear potential. As shown in
Fig.~\ref{figsym}, the symmetry energy  is modified oppositely below
and above the fixed point ($k_F=1.15$fm$^{-1}$), and this is ensured
by the similar modification to the isovector potential. Due to the
strong coupling between protons and neutrons, modifications in the
isoscalar potential are  similar, albeit small. According to the
virial theorem, the modification to the nucleon kinetic energy is
associated with the gradient of the potential. The opposite
modifications to the potential give rise to opposite increments for
the gradient of the potential. Thus, the cancellation between
opposite increments  leads to just a small change to the nucleon
kinetic energy. Eventually, the change in the total binding energy is
small with respect to $\Lambda_{\rm v}$. Nevertheless, since the
empirical binding energies of finite nuclei were firstly reproduced
quite accurately by the best-fit models, the variation of the total
binding energy, albeit small, can not simply be used to constrain the
density dependence of the symmetry energy.

If the variation of the binding energy is expected to be informative
to the density dependence of the symmetry energy, it should be the
relative variation of the binding energies of two systems that
clearly differ by an isovector ingredient. In this work, we are thus
interested in the deexcitation energy of the SSM relative to the
ground state. However, since the isoscalar-isovector coupling
modifies not only the isovector potential but also  the isoscalar
potential,  an efficient probe to the density dependence of the
symmetry energy should be able to separate clearly the modification
to the isovector potential from that to the isoscalar potential. In
the following, it is necessary to exhibit such a separation in
obtaining the deexcitation energy.  In order to analyze the
potentials obtained with the deformed RMF code conveniently in one
dimension, it is necessary to perform the multipole expansion as
\begin{equation}\label{eqme}
U({\bf r})=\sqrt{4\pi}\sum_{L=0,2,4,\cdots} U_L(r)Y_{L0}(\theta),
\end{equation}
where $Y_{L0}$ is the spherical harmonic function. The deexcitation
energy, which is the energy difference between the secondary minimum
(s.m.) and the ground state (g.s.),  is associated with the
corresponding difference of the potentials
\begin{equation}\label{eqpd1}
\Delta U_L^{iso}(r)=U_{L,~\!s.\!m\!.}^{iso}(r)-
U_{L,~\!g\!.\!s\!.}^{iso}(r),
\end{equation}
where the superscript $iso$ represents the isoscalar (IS)  or
isovector (IV). Here, the isoscalar and isovector potentials are
given as $U^{IS}_L(r)=g_\omega\omega(r)-g_\sigma\sigma(r)$ and
$U^{IV}_L(r)=g_\rho b_0(r)$, respectively.

\begin{figure}[thb]
\begin{center}
\vspace*{-15mm}
\includegraphics[height=10.0cm,width=10.0cm]{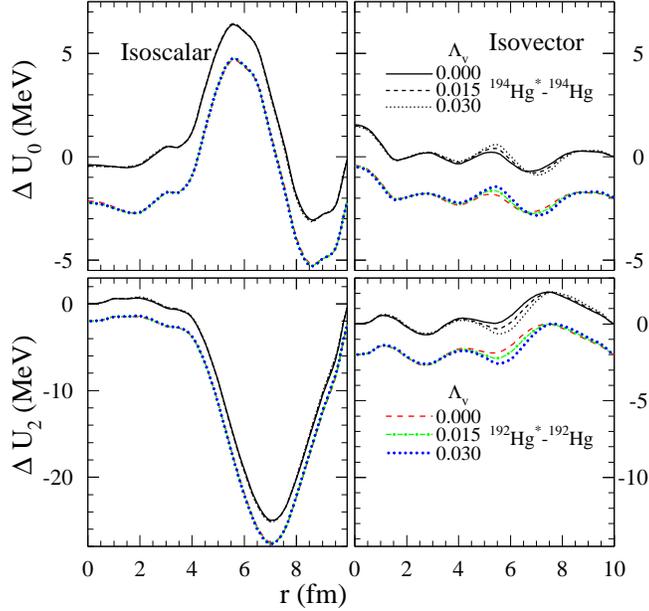}
 \end{center}
\caption{Difference of the potentials between the SSM and ground
state of $^{192,194}$Hg as a function of the radius for various
$\Lambda_{\rm v}$. The calculation is performed within the NL3*. The
left panel is for the isoscalar potential and the right panel for the
isovector one. The result for $^{192}$Hg is displaced downwards by 2
MeV for clarity. \label{fpot}}
\end{figure}

Fig.~\ref{fpot} displays the difference of potentials between the SSM
and  ground state for $^{192,194}$Hg as a function of  radius. In
Fig.~\ref{fpot},  only the two most important components $U_0(r)$ and
$U_2(r)$ that are respectively responsible for the spherical matter
distribution and quadrupole deformation are drawn, while the
difference between the higher-L components of the potentials is
becoming clearly smaller and  much less sensitive to the
$\Lambda_{\rm v}$. It is shown in Fig.~\ref{fpot} that the difference
of the isoscalar potential is almost independent of the
isoscalar-isovector coupling $\Lambda_{\rm v}$. On the other hand,
the difference of the isovector potential varies significantly with
the $\Lambda_{\rm v}$ in the surface region. Especially, a large
variation against the $\Lambda_{\rm v}$ is seen for the isovector
potential difference $\Delta U_2(r)$, associated with the large
oblate-prolate shape difference between the ground state and SSM in
$^{192}$Hg and $^{194}$Hg. These naturally exhibit a clear separation
of the relative variation of the isovector potentials from that of
the isoscalar potentials. In particular, the relative variation of
isoscalar portions of the two states with respect to the
$\Lambda_{\rm v}$ nearly cancels out.

Since the isoscalar-isovector coupling modifies just slightly the
nucleon potentials, it can be treated as a small residue interaction.
This rules out the occurrence of large coherent changes in nuclear
potentials of the ground and metastable states  while with slight
relative changes between various states, and thus the small change in
total binding energies can be determined by that of nuclear
potentials with rather stable nuclear structures. Consequently, the
shift of the deexcitation energy caused by the isoscalar-isovector
coupling can be obtained from the small relative change in  nuclear
potentials between the ground state and SSM. Due to the nearly exact
cancellation between the variations of isoscalar potentials  in the
SSM and ground state, the variation of the deexcitation energy
results predominantly from the modification to the isovector
potential caused by the isoscalar-isovector coupling. Besides from
the potential part, the kinetic energy and the nonlinear term of the
isoscalar-isovector coupling also directly cause the variation of the
deexcitation energy, but they can after all be determined from the
nuclear potential, for instance, according to Eqs.(\ref{eqmeson}) and
(\ref{eqviri}). Eventually, this builds up a direct relationship
between the uncertainty of the deexcitation energy and the
modification to the symmetry energy.

 \begin{table}[bht]
\caption{Properties of the ground state and SSM for $^{194}$Hg within
the NL3* with respect to $\Lambda_{\rm v}$. The binding energy per
nucleon (B/A), charge radius ($r_c$), neutron skin thickness
($r_p-r_n$), and quadrupole deformation parameter $\beta$ are listed.
The properties of the SSM are denoted by the subscript $SD$. Energies
are in unit of MeV and radii in unit of fm. } \label{tab1}
 \begin{center}
    \begin{tabular}{ c c| c c c c c c c }
\hline
 $\Lambda_{\rm v}$&$g_\rho$  & $B/A$ & $r_c$ & $r_n-r_p$ & $\beta$&
(B/A)$_{SD}$&  $(r_n-r_p)_{SD}$& $\beta_{SD}$
 \\ \hline
  0.000  &  4.5748 & 7.913  &   5.460   &  0.213  & -0.146 &  7.885 & 0.182  &   0.625  \\
  0.010  &  4.9005 & 7.928  &   5.461   &  0.189  & -0.145 &  7.898 & 0.159  &   0.622  \\
  0.020  &  5.3074 & 7.940  &   5.464   &  0.165  & -0.144 &  7.907 & 0.137  &   0.618  \\
  0.030  &  5.8360 & 7.948  &   5.468   &  0.141  & -0.142 &  7.914 & 0.115  &   0.614  \\

  \hline
     \end{tabular}
  \end{center}
 \end{table}

In Table~\ref{tab1}, we tabulate as an example calculated quantities
of the ground state and SSM of $^{194}$Hg with respect to the
isoscalar-isovector coupling constant $\Lambda_{\rm v}$.   We see
that except for the neutron skin thickness all the properties of
$^{194}$Hg depend just slightly on the $\Lambda_{\rm v}$. With the
inclusion of the isoscalar-isovector coupling, the neutron skin
thickness reduces appreciably. These are consistent with the early
findings for heavy nuclei in the literature~\cite{ho01,todd}. It is
seen in Table~\ref{tab1} that the SSM has a different neutron skin
thickness from that of the ground state. This is due to the
excitation of protons and neutrons in the vicinity of Fermi surfaces
to different shells. Specifically, in $^{194}$Hg neutrons with the
configuration $2f_{5/2}^2 1h_{9/2}^4$ that are  below the major shell
$N=126$ in the ground state are excited to intruder orbitals that
originate from spherical orbitals $2g_{9/2}1i_{11/2}$ above the major
shell $N=126$. The neutron occupation in intruder orbitals forms the
important configuration $2g_{9/2}^4 1i_{11/2}^2$ for the
superdeformation in the SSM.  For protons, similar excitation to
intruder orbitals occurs from the shell $50<Z<82$ in the ground state
up to the shell $Z>82$ in the SSM. Besides the characteristic
excitation between the major shells, the excitation in subshells also
contributes to the formation of the superdeformation in the SSM.
Thus, the creation of interior holes and occupation of exterior
orbitals  at different major shells for protons and neutrons results
in an appreciably different neutron skin thickness in the SSM from
that in the ground state.  As a result, the SSM differs from the
ground state by an isovector ingredient from which the sensitivity of
the deexcitation energy to the isoscalar-isovector coupling
originates. For other Hg isotopes, the results are similar to those
for $^{194}$Hg, and are not given here.
\begin{table}[htb]
\caption{Deexcitation energies (MeV) of the SSM relative to ground
states for a few Hg isotopes with respect to the $\Lambda_{\rm v}$.
The experimental values for $^{192,194}$Hg and $^{191}$Hg are taken
from Refs.~\cite{lau00,siem04}, respectively. \label{tab2}}
 \begin{center}
    \begin{tabular}{c |c c c c c c}
\hline
    Model&$\Lambda_{\rm v}$& $^{194}$Hg & $^{192}$Hg   & $^{191}$Hg
   \\ \hline
NL3* &0.000  &   5.52 & 4.03  &   3.59   \\
     &0.010  &   5.81 & 4.40  &   3.94   \\
     &0.020  &   6.27 & 4.73  &   4.26   \\
     &0.030  &   6.60 & 5.03  &   4.44   \\
\hline
   Expt.&  &    6.0 &   5.3 & 4.6      \\
  \hline
     \end{tabular}
  \end{center}
 \end{table}

In Table~\ref{tab2}, we list the deexcitation energies relative to
the respective ground state for $^{194}$Hg, $^{192}$Hg and $^{191}$Hg
that have experimental data~\cite{lau00,siem04}. We may observe from
Table~\ref{tab2} and Fig.~\ref{figsym} that the deexcitation energies
are sensitive to differences in the symmetry energy. Comparing with
the experimental data, it is seen that the description for the
deexcitation energies can be largely improved by including the
isoscalar-isovector coupling. With $\Lambda{\rm v}=0.03$, the
deexcitation energies for $^{194}$Hg, $^{192}$Hg and $^{191}$Hg agree
fairly well with the data within acceptable accuracy. This confirms
the softening of the symmetry energy.

\begin{table}[htb]
\caption{Deexcitation energies  for Hg isotopes recalculated with the
slightly modified $m_\sigma$ with respect to the $\Lambda_{\rm v}$.
As compared to $g_\rho$ in Table~\ref{tab1}, it is just slightly
modified to keep the symmetry energy unchanged at
$k_F=1.15$fm$^{-1}$. Ground-state binding energies per nucleon $B/A$
are  listed for $^{194}$Hg. \label{tab3}}
 \begin{center}
    \begin{tabular}{c c c| c c c c c}
\hline $\Lambda_{\rm v}$ & $m_\sigma$ (MeV) & $g_\rho$ & $^{194}$Hg &
$^{192}$Hg & $^{191}$Hg&$B/A$
   \\ \hline
  0.000  &502.5742  & 4.5748   &   5.51  & 4.06  &   3.59   & 7.913 \\
  0.010  &502.6050  & 4.9006   &   5.63  & 4.37  &   3.85   & 7.912 \\
  0.020  &502.6310  & 5.3076   &   6.19  & 4.62  &   4.29   & 7.913 \\
  0.030  &502.6465  & 5.8363   &   6.54  & 4.99  &   4.46   & 7.913 \\
     \hline
     \end{tabular}
  \end{center}
 \end{table}

Now, we clarify the concern whether the variation of deexcitation
energies can be disguised by the much larger change in the
ground-state energy, which is about 7 MeV for the $\Lambda_{\rm v}$
of interest, as estimated in Table~\ref{tab1}. To do this, we just
need to observe the variation of the deexcitation energy in the case
of the constant ground-state energy. It is possible to reduce or even
eliminate the variation of the ground-state binding energy by
suitably choosing the fixed point which can affect the extent of
cancellations, whereas this is numerically complicated.
Alternatively, the elimination of the variation in the ground-state
energy can actually be fulfilled by slightly readjusting the
meson-nucleon coupling constants or meson masses. Without priority,
here we realize it by slightly readjusting the  $\sigma$ meson mass
$m_\sigma$. The recalculated results are given in Table~\ref{tab3}.
As seen in Table~\ref{tab3}, the variation of the total ground-state
binding energy is nearly eliminated by just modifying the $m_\sigma$
up to 0.07 MeV. With this slight readjustment of the $m_\sigma$,  the
modification to the incompressibility is just about 0.8 MeV.
Noticeably, the variation of  deexcitation energies is almost
unchanged, as compared with that given in Table~\ref{tab2}.
Definitely, the sensitivity of deexcitation energies to differences
in the symmetry energy  is irrelevant to the variation of the
ground-state energy. The underlying physics for this is attributed to
the fact that  the variation of the deexcitation energy is
conditioned predominantly on the modification to the isovector
potential that changes the density dependence of the symmetry energy,
as analyzed for results shown in Fig.~\ref{fpot}.

On the other hand, the difference between  deexcitation energies of
$^{194}$Hg and $^{192}$Hg is still large (about 1.5 MeV), as compared
to the experimental value 0.7 MeV, and is not reduced  by including
the isoscalar-isovector coupling. It means that the accurate
deexcitation energies for $^{192}$Hg and $^{194}$Hg both can not be
obtained with the same $\Lambda_{\rm v}$. It is seen in
Table~\ref{tab2} that the difference between the deexcitation
energies for $^{194}$Hg and $^{192}$Hg is not sensitive to the
$\Lambda_{\rm v}$. This is because the isospin asymmetries $\delta$
in $^{194}$Hg and $^{192}$Hg are not very different. Consequently,
the considerable reduction of the difference between the deexcitation
energies for $^{194}$Hg and $^{192}$Hg should resort to the
alteration of the isoscalar potential. For instance, this difference
is reduced to be 0.5 MeV with the RMF parameter set TM1~\cite{tm1}
whose isoscalar potential is moderately different from that with the
NL3* due to the inclusion of the $\omega$-meson self-interaction.
However, with the $\omega$-meson self-interaction, the softened
vector potential leads to small deexcitation energies and much
shallower potential wells of the SSM. As deexcitation energies for
$^{192}$Hg and $^{194}$Hg are both much smaller than the  data within
the TM1,  the isoscalar-isovector coupling can be included to partly
compensate this discrepancy, whereas an overall compensation seems
yet to be available in the model that features the $\omega$-meson
self-interaction.

\begin{figure}[thb]
\begin{center}
\vspace*{-15mm}
\includegraphics[height=10.0cm,width=10.0cm]{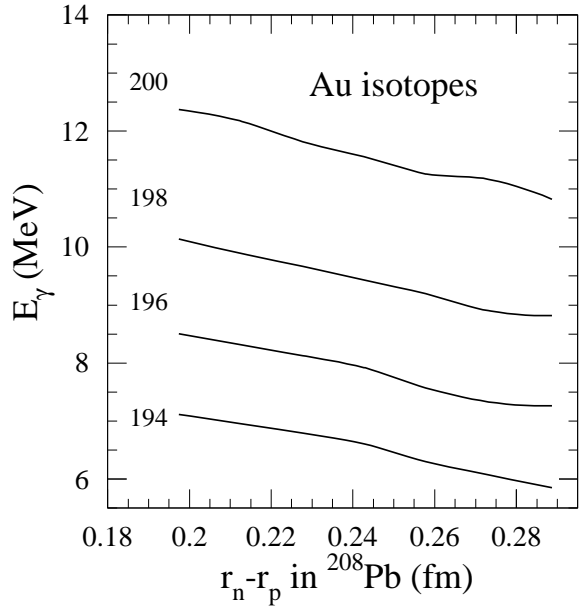}
 \end{center}
\caption{Deexcitation energies  ($E_\gamma$)  of the SSM for odd-odd
Au isotopes versus the neutron skin thickness in $^{208}$Pb with the
NL3*. The results are obtained by varying the $\Lambda_{\rm v}$
properly with $g_\rho$. The mass number is marked for each curve.
\label{figau}}
\end{figure}
Further, we notice that the deexcitation energy is nearly of the same
magnitude as the pairing energy.  The correct deexcitation energy of
some nuclei may be obtained with the inclusion of the
isospin-dependent pairing interactions. For instance, using different
pairing gaps and strengths in Ref.~\cite{la98}, the deexcitation
energy of $^{194}$Hg with the NL3 was reproduced in nice agreement
with the experimental value. On the other hand, the difference of the
deexcitation energies between $^{192}$Hg and $^{194}$Hg was still 0.8
MeV larger than the experimental value~\cite{la98}. If this
prescription is applied to resolve the discrepancy of the
deexcitation energy with the experiments for other isotopes, it will
impose stringent constraints on  pairing strengths. On the other
hand, the symmetry energy in finite nuclei can successfully be
extracted regardless of the pairing interaction~\cite{ban06}. This
implies that the observable response to the symmetry energy would be
rather independent of the pairing interaction.  Specifically, we find
that the deexcitation energies of the SSM in odd-odd nuclei in the
region $A\sim190$ just depend very weakly on the pairing interaction,
while the sensitive dependence on the isoscalar-isovector coupling
does not alter at all. The former is because the pairing interaction
in odd-odd nuclei is substantially suppressed by the Pauli blocking
and the nucleonic current that breaks the double degeneracy. If  the
pairing interaction turns off while reviving the nucleonic current
and the contribution of the $\pi$ meson in these odd-odd
nuclei~\cite{jr05}, the deexcitation energy is just changed by about
0.2 MeV. Therefore, no appreciable effect on deexcitation energies
can be observed by modifying pairing strengths in these odd-odd
nuclei. In Fig.~\ref{figau}, we display as examples  the deexcitation
energies for a few odd-odd Au isotopes  as a function of the neutron
skin thickness for $^{208}$Pb. Since the neutron skin thickness in
$^{208}$Pb depends almost linearly on the $\Lambda_{\rm
v}$~\cite{ho01,todd}, the sensitive dependence on the $\Lambda_{\rm
v}$ can be observed for  deexcitation energies of odd-odd Au
isotopes. We can thus expect that the measurement of the deexcitation
energy of these odd-odd isotopes can be useful not only for
constraining the density dependence of the symmetry energy but also
clarifying this theoretical uncertainty.

Next, let us discuss more the factors that are necessary for the
description of the deexcitation energy. First of all, since the SSM
is a result of the collective excitation that undergoes a significant
isovector change, the appropriate isovector potential of the model is
necessary. We find from Table~\ref{tab2} and \ref{tab3} that the
description for deexcitation energies of various isotopes is
improved by including the isoscalar-isovector coupling. In other
words, the inclusion of the isoscalar-isovector coupling improves the
isovector potential of the model. Secondly, we note that the
isoscalar potential is also an important ingredient to carry out
correct deexcitation energies. Especially, the difference between the
deexcitation energies for the neighboring even-even isotopes is
tightly associated with the appropriate fit to the isoscalar
potential. The present study indicates that an overall description
for the deexcitation energies and difference between them needs to
reconstruct the  model appropriately with the isovector potential
that brings out a softened symmetry energy.

At last, we should stress that though the present work is performed
based on a specific parametrization (NL3*), the conclusion that the
deexcitation energy is sensitive to differences in the symmetry
energy is rather free of this specific RMF model. We have examined
this with many other RMF models. The current situation is that most
best-fit models have  been constructed regardless of the detail of
the density dependence of the symmetry energy. Thus, the agreement
with data of deexcitation energies may play a significant role in
reconstructing models with the appropriate density dependence of the
symmetry energy. Moreover, we have noted that the treatment of the
nucleon pairing with the BCS theory is comparatively simple as
compared with the Bogoliubov approach. This would affect the accuracy
of the correlation between deexcitation energies of even nuclei and
the density dependence of the symmetry energy.  One can expect the
improvement with the relativistic Hartree-Bogoliubov models. However,
this is beyond the capacity and scope of this preliminary effort.

\section{Summary}
\label{summary}

In summary, we have studied the relationship between the deexcitation
energies of the SSM in heavy nuclei and the density dependence of the
symmetry energy. The investigation is based on the RMF model (NL3*)
with the isoscalar-isovector coupling included to modify the density
dependence of the symmetry energy. It is found that the uncertainty
of the deexcitation energy originates almost uniquely from the
modification to the isovector potential induced by the
isoscalar-isovector coupling. The deexcitation energies can thus
serve as a theoretical probe to the density dependence of the
symmetry energy. As a result, we find that the theoretical estimates
of the deexcitation energies of Hg isotopes can be improved by the
inclusion of the isoscalar-isovector coupling that softens the
symmetry energy. To separate the effect of the symmetry energy on
deexcitation energies from that of pairing correlations,  we propose
to measure the deexcitation energies of odd-odd heavy nuclei in the
region $A\sim190$ such as Au isotopes that are nearly independent of
pairing correlations. This can be used to constrain the isovector
content of the model and thus the density dependence of the symmetry
energy.

\section*{Acknowledgement}

One of authors (W.Z.Jiang) thanks Prof. Bao-An Li for the enlightening
discussions. Authors thank the partial support from the Kavli Institute for
Theoretical Physics China. The work is also supported in part by the NNSF of
China under Grant Nos. 10975033, 10535010 and 10675090, the Jiangsu NSF under
Grant No. BK2009261,  the KI Project of the CAS under Grant No. KJXC3-SYW-N2,
and the China Major State BRD Program under Contract No. 2007CB815004.


\begin{thebibliography}{99}
\bibitem{lat01} {\small J. M. Lattimer and M. Prakash, Phys. Rep. \textbf{333},
121 (2000); Astrophys. J. \textbf{550}, 426 (2001); Science
\textbf{304}, 536 (2004).}

\bibitem{ho01} C. J. Horowitz, J. Piekarewicz, Phys. Rev. Lett. \textbf{86},
5647 (2001); \textbf{64}, 062802(R)(2001).

\bibitem{ho02}  C. J. Horowitz, J. Piekarewicz, Phys. Rev. C \textbf{66},055803 (2002).

\bibitem{steiner05a} {\ A. W. Steiner, M. Prakash, J. M. Lattimer and P. J.
Ellis, Phys. Rep. \textbf{411}, 325 (2005).}

\bibitem{todd}B. G. Todd, J. Piekarewicz, Phys. Rev.
C \textbf{67}, 044317 (2003).

\bibitem{ji05} W. Z. Jiang, Y. L. Zhao, Phys. Lett. B \textbf{617},33 (2005).

\bibitem{ba08} B. A. Li, L. W. Chen, and C. M. Ko, Phys. Rep. \textbf{464}, 113 (2008).

\bibitem{ts04} M. B. Tsang et.al. Phys. Rev. Lett. \textbf{92}, 062701 (2004).

\bibitem{ch05} L. W. Chen, C. M. Ko, B. A. Li, Phys. Rev. Lett. \textbf{94},
032701 (2005).

\bibitem{li05} B. A. Li, L. W. Chen, Phys. Rev. C \textbf{72}, 064611 (2005).

\bibitem{Fopi07}W. Reisdorf et al. (FOPI Collaboration), Nucl. Phys. A \textbf{781}, 459 (2007).

\bibitem{xiao08} Z. G. Xiao, B. A. Li, L. W. Chen, G. C. Yong and M.
Zhang, Phys. Rev. Lett. \textbf{102}, 062502 (2009).

\bibitem{piek06} J. Piekarewicz, and S. P. Weppner, Nucl. Phys. A \textbf{778},
10 (2006).

\bibitem{jeff}R. Michaels, P. A. Souder, G. M. Urciuoli, spokespersons,
Jefferson Laboratory Experiment E-00-003.
\bibitem{ma86} W. C. Ma, A. V. Ramayya, J. H. Hamilton, et.al., Phys. Lett. B
\textbf{167}, 277 (1986).

\bibitem{dr88} G. D. Dracoulis, A. E. Stuchbery, A. O. Macchiavelli, et al., Phys.
Lett. B \textbf{208}, 365 (1988).

\bibitem{wo92} J. L. Wood, K. Heyde, W. Nazarewicz, M. Huyse, and P. Van Duppen,
Phys. Rep. \textbf{215}, 101 (1992).
\bibitem{he98}P.-H. Heenen, J. Dobaczewski, W. Nazarewicz, P. Bonche, and T. L.
Khoo, Phys. Rev. C \textbf{57}, 1719 (1998).

\bibitem{lau00} T. Lauritsen, T. L. Khoo, I. Ahmad, et.al., Phys. Rev. C \textbf{62},
044316 (2000).

\bibitem{vr03}D. Vretenar, T. Niksic, and P. Ring, Phys. Rev. C \textbf{68}, 024310
(2003).

\bibitem{pa05}N. Paar, T.Niksic, D. Vretenar, and P. Ring, Phys.
Lett. B \textbf{606}, 288 (2005).

\bibitem{pi06}J. Piekarewicz, Phys. Rev. C \textbf{73}, 044325
(2006).

\bibitem{liang07}J. Liang, L. G. Cao, and  Z. Y. Ma,
Phys. Rev. C \textbf{75}, 054320 (2007).

\bibitem{ma02}Z. Y. Ma, L. Liu, Phys. Rev. C \textbf{66}, 024321 (2002).


\bibitem{ji07} W. Z. Jiang, B. A. Li and L. W. Chen, Phys. Lett. B \textbf{653},
184 (2007).

\bibitem{mn2}P. M\"oller and J. R. Nix, Nucl. Phys. A \textbf{536}, 20 (1992).

\bibitem{ri96}P. Ring, Prog. Part. Nucl. Phys. \textbf{37}, 193 (1996).

\bibitem{meng06}J. Meng, H. Toki, S. G. Zhou, S. Q. Zhang, W. H. Long and L. S. Geng,
Prog. Part. Nucl. Phys. \textbf{57}, 470 (2006).

\bibitem{ga90}Y. K. Gambhir, P. Ring and A. Thimet, Ann. Phys. (N.Y.)
\textbf{198}, 132 (1990).

\bibitem{nl3new}G. A. Lalazissis, S. Karatzikos, R. Fossion, D. P. Arteaga,
A. V. Afanasjev, and P. Ring, Phys. Lett. B \textbf{671}, 36 (2009).

\bibitem{nl3} G. A. Lalazissis, J. K\"onig, and P. Ring, Phys. Rev.
C \textbf{55}, 540 (1997).

\bibitem{se86}B. D. Serot and J. D. Walecka, Adv. Nucl.
Phys. \textbf{16}, 1(1986).

\bibitem{siem04} S. Siem, P. Reiter,  T. L. Khoo,
et.al., Phys. Rev. C \textbf{70}, 014303 (2004).

\bibitem{tm1} Y. Sugahara and H. Toki, Nucl. Phys. A \textbf{579}, 557 (1994).

\bibitem{la98}G. A. Lalazissis and P. Ring, Phys. Lett. B \textbf{427}, 225 (1998).

\bibitem{ban06}S. Ban, J. Meng, W. Satula, and R. A. Wyss, Phys.
Lett. B \textbf{633}, 231 (2006).

\bibitem{jr05}W. Z. Jiang, Z. Z. Ren, T. T. Wang, Y. L. Zhao and Z. Y.
Zhu, Eur. Phys. J. A \textbf{25}, 29 (2005).


\end{thebibliography}
\end{document}